# Exchange biasing and electric polarization with YMnO$_3$

*X. Martí, F. Sánchez, D. Hrabovsky, L. Fàbrega, A. Ruyter, J. Fontcuberta*

Institut de Ciència de Materials de Barcelona - CSIC, Campus U.A.B., Bellaterra 08193, Spain

*V. Laukhin*

Institut de Ciència de Materials de Barcelona - CSIC, Campus U.A.B., Bellaterra 08193, and Institut Català de Recerca i Estudis Avançats (ICREA), Barcelona, Spain

*V. Skumryev*

Departament de Física, Universitat Autònoma de Barcelona, Bellaterra 08193, and Institut Català de Recerca i Estudis Avançats (ICREA), Barcelona, Spain

*M.V. García-Cuenca, C. Ferrater, M. Varela*

Departament de Física Aplicada i Òptica, Universitat de Barcelona, Diagonal 647, Barcelona 08028, Spain

*A. Vilà*

Departament d'Electrònica. Universitat de Barcelona, Diagonal 647, Barcelona 08028, Spain

*U. Lüders, J.F. Bobo*

LNMH CNRS-ONERA, BP 4025 31055 Toulouse Cedex 4, France

**Abstract**

We report on the growth and functional characterization of epitaxial thin films of the multiferroic YMnO$_3$. We show that using Pt as a seed layer on SrTiO$_3$(111) substrates, epitaxial YMnO$_3$ films (0001) textured are obtained. An atomic force microscope has been used to polarize electric domains revealing the ferroelectric nature of the film. When a Permalloy layer is grown on top of the YMnO$_3$(0001) film, clear indications of exchange bias and enhanced coercivity are observed at low temperature. The observation of coexisting antiferromagnetism and electrical polarization suggests that the biferroic character of YMnO$_3$ can be exploited in novel devices.

Multiferroic materials, displaying simultaneously the occurrence of magnetic and ferroelectric orders, are nowadays of the highest interest because of possible applications where dual tuning of the magnetic and electric response could be of relevance [1,2]. Whereas much effort is being focused to the scarce materials displaying ferromagnetic and ferroelectric response [3], the functional exploitation of the more abundant materials displaying antiferromagnetic (AF) and ferroelectric (FE) orders has received less attention. Hexagonal $YMnO_3$ (YMO) belongs to this class of materials. It displays a high temperature FE transition (at about 900 ºC) and a low temperature AF transition ($T_N \approx$ 70 K). Possible use of the FE properties of YMO in ferroelectric devices has been already highlighted [4] and YMO thin films have been subsequently grown with a variety of techniques [4,5]. Electrical polarization in YMO is along the hexagonal c-axis [6] whereas the magnetic moments of $Mn^{3+}$ lie in the perpendicular plane, forming a triangular, geometrically frustrated, network of AF coupled spins [7,8].

Antiferromagnets are used in spintronics by exploiting the exchange bias (EB) with suitable ferromagnets. Therefore, YMO could be also used as pinning layer if significant EB could be stimulated across an interface with a ferromagnet. Indeed, it has been recently proposed that YMO displays EB when grown textured along the [11$\underline{2}$1] direction but almost no exchange bias was observed for [0001] films [9]. As the [0001] direction is the polarization direction it seemed that prospectives of YMO for multifunctional devices were rather limited. In this paper we will show that YMO films with a unique [0001]-axis texture can be grown on (111)-oriented $SrTiO_3$ (STO) single crystalline substrates when buffered with a thin Pt layer and that these films display both ferroelectric character and a substantial EB. As FE and AF domains are coupled in YMO [10] it turns out that YMO could be exploited for multifunctional operation.

Pt films were deposited on a STO(111) substrate by dc sputtering. Previous to the deposition, substrates were annealed in-situ during 5 min at 800 ºC and, afterwards, Pt (~8 nm thick) was deposited at $5·10^{-3}$ mbar pure Ar and at substrate temperature of 500 ºC. The Pt buffered substrate was transferred to a pulsed laser deposition chamber and the YMO was subsequently grown, using a stoichiometric $YMnO_3$ target. Films were deposited at a substrate temperature of 800 ºC in 0.2 mbar of oxygen pressure. At the end of the growth the samples were cooled down and 1 atm of oxygen was introduced in the chamber at 500 ºC. A Permalloy ($Ni_{81}Fe_{19}$, Py) layer, 15 nm thick, was subsequently grown by dc sputtering at room temperature, under a magnetic field of 100 Oe, in order to build the Py/YMO/Pt//STO(111) heterostructure.

The crystal structure was investigated by X-ray diffractometry (XRD) using Cu $K_\alpha$ radiation. Pt film thickness was determined by modeling Laue fringes observed in XRD patterns, whereas profilometry was used in other cases. Magnetic and electric transport properties were measured by using a superconducting quantum interference device and a PPMS from Quantum Design. Piezoresponse force microscopy (PFM) with a Nanoscope Dimension 3100 microscope (Digital Instruments) was used to measure the piezoelectric response. The bottom Pt electrode was connected to ground by using silver paste, while the voltages $V_{DC}$, $V_{AC}$, to polarize the sample and characterize the response respectively, were applied to the upper surface of the YMO film through the microscope tip.

In Fig. 1a we show a XRD θ/2θ scan corresponding to a Py/YMO(90nm)/Pt//STO(111) sample. Diffraction peaks from the substrate and the Pt and YMO layers are found. Inspection of the Pt reflections indicates that the Pt film has a (111) texture and the observation of the associated Laue fringes reflects the flatness and crystal quality of the Pt films. The reflections (000$l$) of the hexagonal YMO phase are well visible. Interestingly, no traces of orthorhombic phase are found in spite of being the stable phase when YMO grows directly on STO(111) substrates [11]. Out-of-plane lattice distances were evaluated from Bragg's peaks positions. The out-of-plane parameter of Pt is $d_{(111)} = 2.27$Å, almost coincident with that of bulk Pt whereas the out-of-plane $d_{(0001)} \equiv c$ parameter of YMO is 11.45Å, somewhat expanded with respect to $c_{bulk} = 11.40$Å; the strain is $\varepsilon_{(0001)} = 0.44\%$, suggesting an in-plane compression. Indeed, the in-plane parameter of the YMO film, determined from XRD ϕ-scans, matches that of Pt thus reflecting the epitaxial compressive stress on the YMO films. In Fig. 1b we have collected the pole figures of the STO(110), Pt(11$\underline{1}$), and YMO(11$\underline{2}$4) reflections measured along the pole [111] STO direction. Considering the poles of the Pt(11$\underline{1}$) reflections, one notices the existence of two sets of peaks (labeled A and B) corresponding to the family of planes {11$\underline{1}$} and {$\underline{11}$1} respectively. It indicates that the Pt epitaxial film has grown in two domains, 180º in-plane rotated; the epitaxial relationships with the substrate are [11$\underline{2}$]Pt(111)/[11$\underline{2}$]STO(111) and [$\underline{11}$2]Pt(111)/[11$\underline{2}$]STO(111). The pole figure of the YMO(11$\underline{2}$4) reflections shows six peaks as expected due to the P6$_3$cm symmetry of YMO. Notice that this symmetry does not allow to discern the presence of one or two domains (60º in-plane rotated). However, since YMO grows on bi-domained Pt layer, two domains also in the YMO layer are expected. The corresponding epitaxial relationships are [10$\underline{1}$0]YMO/[11$\underline{2}$]Pt and [10$\underline{1}$0]YMO/[$\underline{11}$2]Pt.

The magnetic hysteresis loops, measured at 100 K and 2 K after a magnetic field (3 kOe) cooling procedure, are displayed in Fig. 2. The loop measured at 100 K is symmetric with coercivity ($H_C$) of only ~10 Oe, very similar to the loops measured at higher temperature. In contrast, the loop measured at 2 K is asymmetric, revealing that an exchange bias ($H_{EB} \approx 61$Oe) accompanied by a significant coercivity enhancement ($H_C \approx 98 Oe$) develops at low temperatures. Fig. 2 includes two magnetization loops measured consecutively (denoted 1st and 2nd). It can be appreciated that there is a significant reduction of $H_{EB}$ upon cycling the field. This training effect [12] is characteristic of AF systems with low or multiaxial anisotropy [13], which likely contributes to the fast decay of $H_{EB}$ upon heating. The temperature dependence of $H_{EB}$ and $H_C$ are shown in Fig. 2 (inset). We notice that $H_{EB}$ decays rapidly when increasing the temperature and vanishs at ~10 K.

The existence of EB also provokes changes of the anisotropic magnetoresistance (AMR) of the upper-lying Py layer [14,15]. AMR was measured by rotating the magnetic field in the film plane while recording the film resistance R as a function of the angle θ between the measuring current and the applied magnetic field ($H_a$). In Fig. 3a we include the R(θ) cycles recorded at several temperatures (100 K, 50 K and 2 K), above and below $T_N$ (~70 K). Prior to any measurement, the sample was field cooled (3 kOe) from room temperature to the measuring temperature and then the field was reduced to the selected $H_a$ value. Data included in Fig. 3 were obtained when the cooling field was fixed at 45º from the current direction and $H_a$ = 40 Oe. We note that this value is smaller that the largest $H_c$ value determined from the magnetization loops (Fig. 2).

At T = 100 K, R(θ) displays the common $\cos^2\theta$ dependence expected for a system where the magnetization follows the magnetic field (Fig. 3a). However, at 50K, R(θ) displays a clear departure of the $\cos^2\theta$ dependence due to the presence of the $H_{EB}$, which becomes more prominent when cooling down (Fig. 3b). In fact, at 2 K the presence of an $H_{EB}$ has dramatically modified the magnetization response when $H_a$ is rotated (Fig. 3c). The strong departure of the R(θ) curve from the $\cos^2\theta$ dependence indicates that, at 2K, the $H_{EB}$ should be much larger than the measuring field (40 Oe), and then larger than the values extracted from the magnetization loops. Observation of larger $H_{EB}$ values in transport measurements results from the different way in which the presence of an $H_{EB}$ is revealed [14]. Notice that in transport measurements exchange bias is monitored via its impact on the magnetization when rotating $H_a$ but there is no need to switch the magnetization as in the common magnetic measurements and thus training effects are less important.

We turn now to the ferroelectric response. First, a background was set over a 2x2 μm² square by scanning it with $V_{DC}$ = +6 V, in contact mode using a tip velocity of 0.8 μm/s. Inside, a smaller square (1x1μm²) was polarized with $V_{DC}$ = -6V, using the same tip velocity. Finally, the PFM response is recorded by using an AC signal ($V_{AC}$ = 3.5V), and scanning over the outer square with a tip velocity of 0.4 μm/s.

In Fig. 4 we show the PFM phase image of a YMO(150nm)/Pt//STO sample. As expected, it displays contrast over the polarized square (bright area, 1x1 μm² at the center), due to the different phase of the PFM response for the up and down domains; although the observed phase-contrast is rather small (~3º), the obvious change of contrast confirms the piezoresponse of the film. Traces of the substrate-induced steps are well visible in the PFM image (as well as in the topographic image –not show-). We stress that contrast in the PFM image remains roughly constant for several hour after poling, thus suggesting a low leakage.

In summary, we have shown that YMO films can be used to induce exchange bias on suitable ferromagnetic layers while preserving the electric polarization. As magnetic and ferroelectric domains in YMO are coupled [10] our results strongly suggest that multiferroic materials with antiferromagnetic and ferroelectric orders can be successfully used to modify the exchange bias, and subsequently the transport and magnetic properties of a ferromagnetic layer, by electric fields.

Financial support by the MEC of the Spanish Government (projects NAN2004-9094 and MAT2005-5656), and FEDER and the STREP project (Nº 033221) of the E.U. are acknowledged. A.R. (permanent address: LEMA Laboratory, UMR 6157 CNRS-CEA, and Faculté des Sciences et Techniques, University of Tours, France) acknowledges Generalitat de Catalunya for a grant. We are thankful to J.M. Pérez for technical support.

**References**

1. M. Fiebig, J. Phys. D: Appl. Phys. **38**, R123 (2005)

2. Ch. Binek and B. Doudin, J. Phys.: Conds. Matter **17**, L39 (2005)

3. N.A. Hill, J. Phys. Chem. B **104**, 6694 (2000)

4. S. Imada, T. Kurakova, E. Tokumitsu, and H. Ishiwara, Jpn. J. Appl. Phys. **40**, 666 (2001); A. Posadas, J.B. Yau, C.H. Ahn, J. Han, S. Gariglio, K. Johnston, K.M. Rabe, and J.B. Neaton, Appl. Phys. Lett. **87**, 171915 (2005); D. Ito, N. Fujimura, T. Yoshimura, and T. Ito, J. Appl. Phys. **93**, 5563 (2003)

5. J. Dho, C.W. Leung, J.L. Driscoll, and M.G. Blamire, J. Cryst. Growth **267**, 548 (2004)

6. see for instance, B.B. van Aken, T.T.M. Palstra, A. Filippetti, and N.A. Spaldin, Nature Mat. **3**, 164 (2004)

7. A, Muñoz, J.A. Alonso, M.J. Martínez-Lope, M.T. Casáis, J.L. Martínez, and M.T. Fernández-Díaz, Phys. Rev. B **62**, 9498 (2000)

8. M. Fiebig, D. Frölich, K. Kohn, St. Leute, Th. Lottermoser, V.V. Pavlov, and R.V. Pisarev, Phys. Rev. Lett. **84**, 5620 (2000)

9. J. Dho and M. G. Blamire, Appl. Phys. Lett. 87, 252504 (2005)

10. M. Fiebig, T. Lottermoser, D. Frohlich, A.V. Goltsev, and R.V. Pisarev, Nature **419**, 818 (2002)

11. X. Martí, F. Sánchez, J. Fontcuberta, M.V. García-Cuenca, C. Ferrater, and M. Varela, J. Appl. Phys. **99**, 08P302 (2006)

12. J. Nogués and I.K. Schuller, J. Magn. Magn. Mat. **192**, 203 (1999)

13. A. Hoffmann, Phys. Rev. Lett. 93, 097203 (2004)

14. B.H. Miller and E.Dan Dahlberg, Appl. Phys. Lett. **69**, 3932 (1996)

15. H. L. Brown, E. Dan Dahlberg, M. Kief, and Ch. Hou, J. Appl. Phys. **91**, 7415 (2002)

**Figure captions**

**Figure 1**: (a) XRD θ/2θ scan of Py/YMO(0001)/Pt(111)//STO(111). Dashed line indicates the angular position where the orthorhombic peaks should appear. (b) Pole figure collecting the STO(110), YMO(११$\underline{2}$4), and Pt(११$\underline{1}$) reflections. The two families of {११$\underline{1}$} and {$\underline{1}$११} planes of Pt are labeled as A and B, respectively.

**Figure 2** Magnetization loops *vs* applied magnetic field, measured at 100 K (open symbols) and 2 K (close symbols), after cooling the sample from 150 K in a field of 3kOe. Two consecutive loops recorded at 2 K are displayed. Inset: temperature dependence of the exchange bias (closed squares) and the coercivity (open circles) fields.

**Figure 3**: Angular dependence of the magnetoresistance (AMR) of the Py film at (a) 100 K, (b) 50 K and (c) 2 K. Measurements were done using a magnetic field of 40 Oe, and after field-cooling (3 kOe) applied at 45º with respect to the measuring current.

**Figure 4** Piezoresponse image of YMnO/Pt//STO(111) after poling an area of 2x2μm² with +6V and repolarizing an areas of 1x1μm² with -6V.

Figure 1

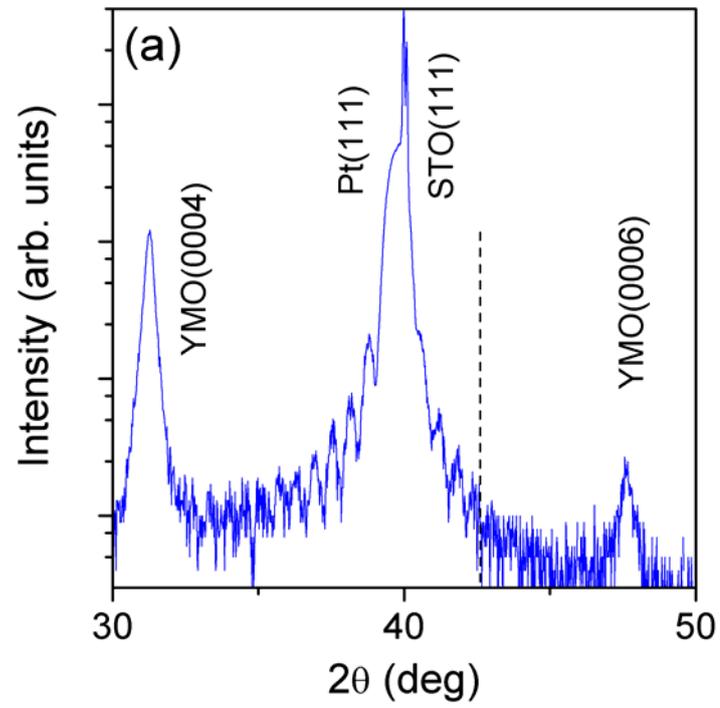

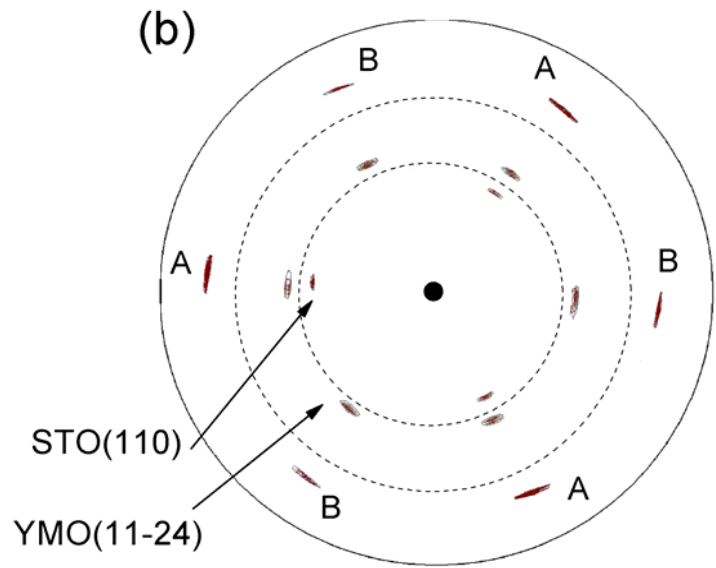

Figure 2

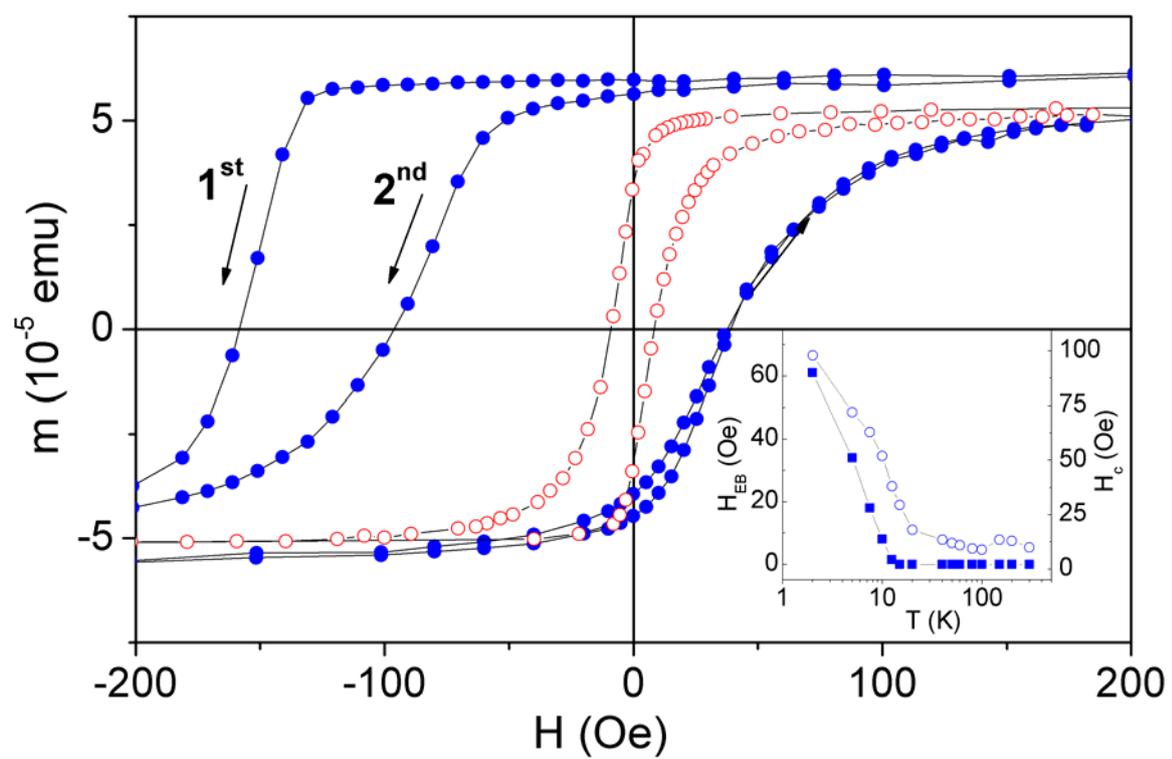

Figure 3

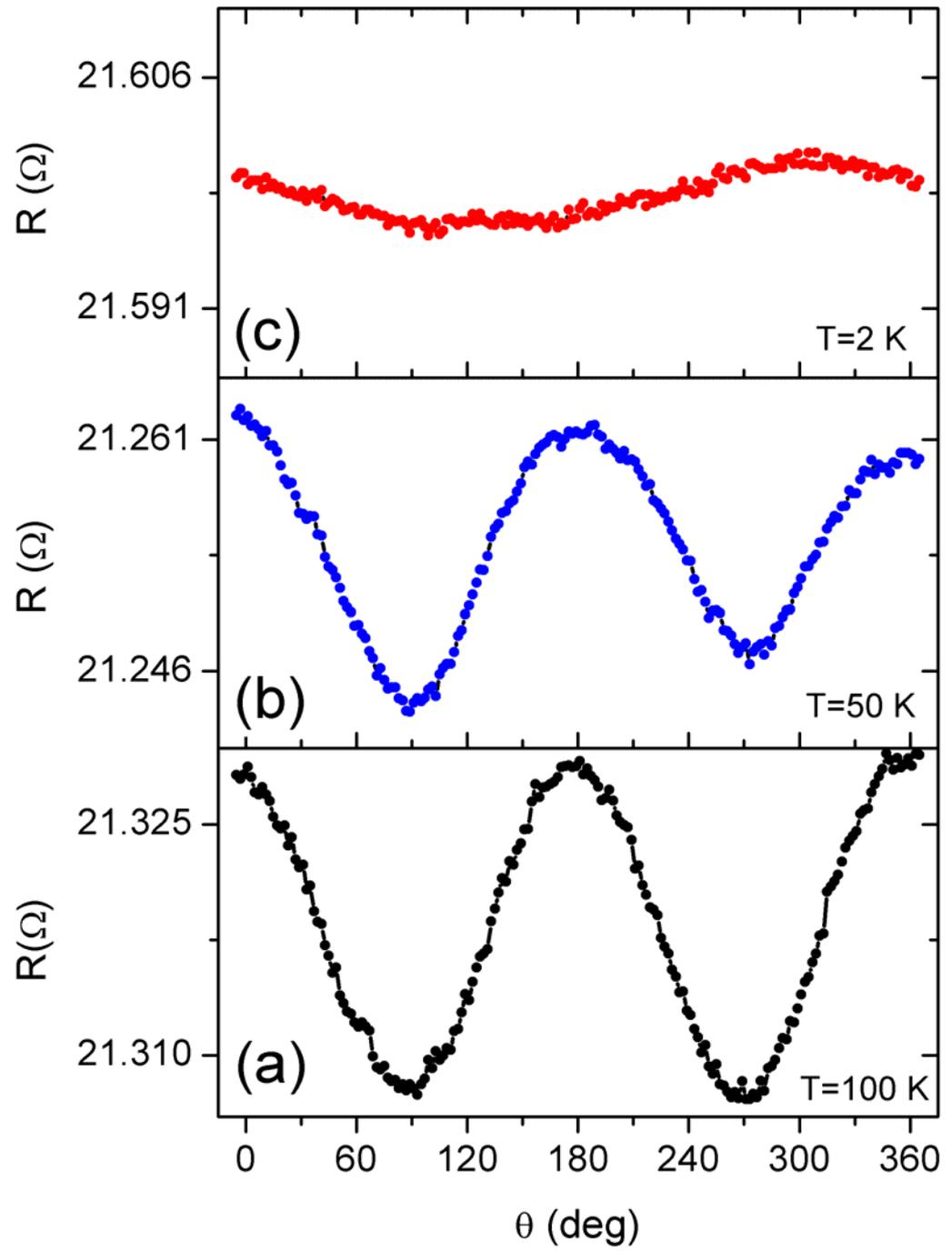

Figure 4

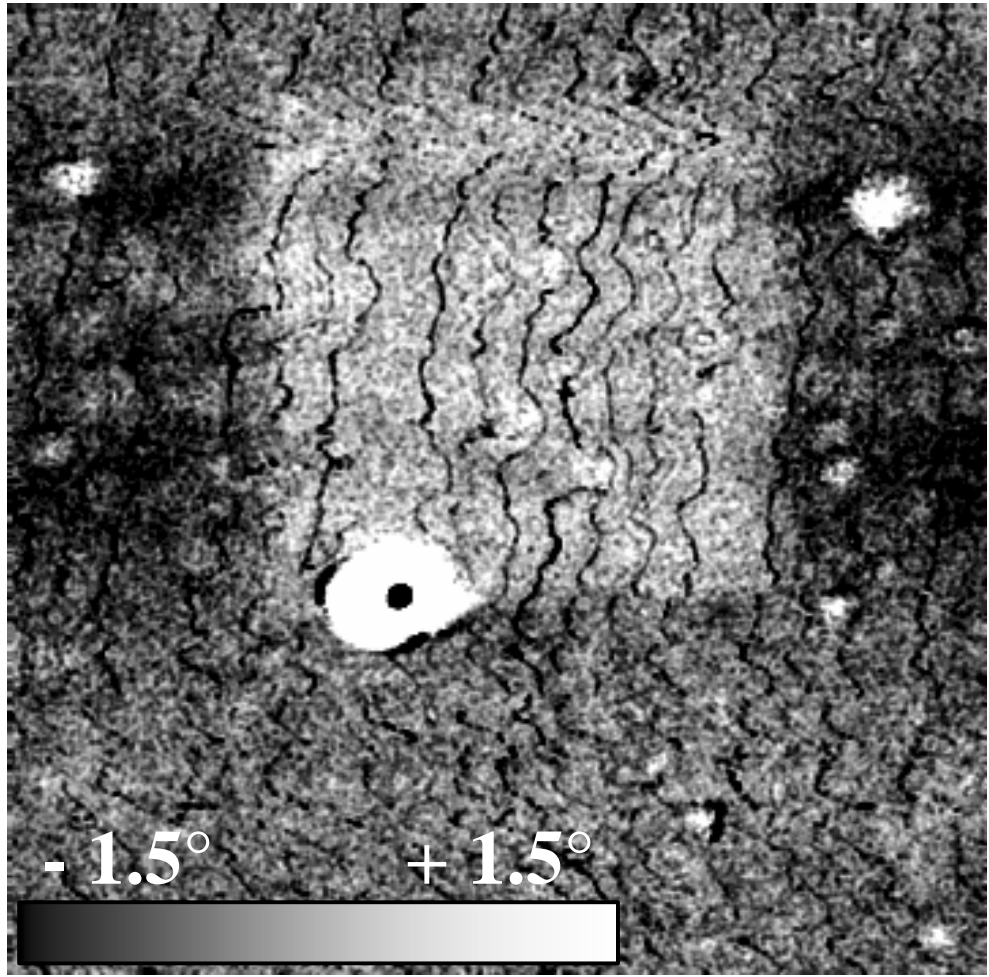